\documentclass[12pt,a4paper]{iopart}
\usepackage{bm}           

\usepackage[pdftex,bookmarks,colorlinks,breaklinks]{hyperref}  

\usepackage{iopams}
\usepackage{graphicx}

\expandafter\let\csname equation*\endcsname\relax 
\expandafter\let\csname endequation*\endcsname\relax 
\usepackage{amsfonts}
\usepackage{amsmath}
\usepackage{bm}

\usepackage{cite}

\usepackage{graphicx}
\usepackage{mathrsfs}

\newcommand{\be}{\begin{equation}}
\newcommand{\ee}{\end{equation}}
\newcommand{\ba}{\begin{eqnarray}}
\newcommand{\ea}{\end{eqnarray}}

\newcommand{\sfrac}[2]{{\textstyle\frac{#1}{#2}}}

\newcommand{\forget}[1]{\iffalse#1\fi}
\newcommand{\forgetmenot}[1]{\iftrue#1\fi}

\newcommand{\del}{\nabla}

\newcommand{\Div}{\hskip0.9pt{\mathrm{div}\hskip2pt}}

\renewcommand{\div}{\Div}

\newcommand{\Curl}{\hskip0.9pt{\mathrm{curl}\hskip2pt}}
\newcommand{\curl}{\Curl}

\newcommand{\Dis}{\hskip0.9pt{\mathrm{dis}\hskip2pt}}
\newcommand{\dis}{\Dis}
\newcommand{\Del}{\hskip0.5pt{\bf \mathrm{D}\hskip0.5pt}}
\newcommand{\sdel}{\Del}
\newcommand{\D}{\Del}

\newcommand{\scal}{\hskip0.5pt\mathcal{S}\hskip0.5pt}
\renewcommand{\vec}{\hskip0.5pt\mathcal{V}\hskip0.5pt}
\newcommand{\ten}{\hskip0.5pt\mathcal{T}\hskip0.5pt}

\renewcommand{\>}{\rangle}

\begin{document}

\title
{Local gauge-invariance at any order in cosmological perturbation theory}

\author{Chris Clarkson}

\address{Astrophysics, Cosmology \& Gravity Centre, and, Department of Mathematics, and,
Applied Mathematics, University of Cape Town, Rondebosch 7701, Cape Town, South Africa}

\ead{\mailto{chris.clarkson@uct.ac.za}}

\begin{abstract}

The relativistic theory of structure formation in cosmology is based mainly on linear perturbations about a homogeneous background. But we are now driven to understand the theory of higher-order perturbations in full detail, both from  observational and theoretical points of view.
An important aspect of this lies in
defining gauge-invariant perturbations at any order. We present a new covariant approach to this based on a local separation of tensors into their scalar, vector and tensor parts. Such a local decomposition necessarily requires a trio of mutually annihilating differential operators which form the basis for defining gauge-invariant objects. It makes no use of non-local Green's functions or pre-defined gauges, and can be used to define families of scalar, vector and tensor modes at any order one chooses. 

\end{abstract}


\section{Introduction}

A challenging problem in relativistic cosmology is perturbation theory around a homogeneous and isotropic background when extended beyond first-order. A particular difficulty lies in preserving general covariance under perturbations, which necessitates considering general coordinate transformations order-by-order alongside a pertubative expansion, and trying to eliminate gauge degrees of freedom. For a tensor to be gauge-invariant (GI) at a given order, it must vanish (or be constant) at all lower orders~(see \cite{Bruni:1996im} for a full discussion). 
It is widely acknowledged that using tensors which are fully gauge invariant is the best way to construct higher-order perturbation theory. 
One important result of Nakamura's~\cite{Nakamura:2010yg} has been to show how starting from gauge-invariant first-order quantities in the metric approach one can generate GI objects at second- and higher-order. Thus one can define a second-order GI tensor perturbation of the metric in the Poisson gauge, for example, but this perturbation is gauge-dependent in the sense that it depends on using the Poisson gauge to begin with; starting in a different gauge one has a different tensor perturbation, even though it too can be GI (see~\cite{Malik:2008im} for details).

An important restriction for these gauge-based GI quantities is that they are defined non-locally, which requires assumptions about unknowable boundary conditions at infinity and inherently relies on assumptions about conditions outside our horizon. While this is fine in a mathematical sense, it is at the very least peculiar in cosmology because such conditions really are, in a fundamental sense, unknowable, and cannot influence conditions within our horizon.

In addition to this, it is often not clear what the GI quantities mean in the following sense: consider a second-order GI extension to the usual Newtonian potential; because this is GI it must correspond to the leading term of some tensor which vanishes at first-order (since we know GI variables are tensors which vanish at lower orders), but we do not know what this object is nor how to find it. 

All constructions of higher-order perturbation theory so far are based on the so-called metric approach to perturbation theory, whereby the metric is split into various parts whose dynamics is governed by different components of the field equations. An important alternative approach to linear perturbation theory has been the 1+3 covariant approach, which has gauge-invariance built in~\cite{Ellis:1998ct,Tsagas:2007yx}. But this approach has languished at first-order, unable to compete with the metric approach because of the difficulty of defining GI variables at higher-order, except in special circumstances. This problem we rectify here by providing a new general technique for defining GI objects at any order covariantly~-- and these quantities can be used in the metric approach too.

This method offers the following advantages: it is conceptually simple, being based on permutations of three mutually annihilating differential operators; it is local as there are no Green's functions of $\nabla^2+nK$ involved; it is covariant, so no gauges or coordinates are necessary to define these quantities; and finally, variables are defined such that they are automatically scalar, vector or tensor modes.

\section{The covariant approach to cosmological perturbations}

In cosmology there exists physically defined observers and therefore reference frames with which to consider physical quantities. Any observable quantity necessarily relies on such an observer being chosen. This underlies the 1+3 covariant approach to GR initiated by Hawking, Ehlers and Ellis~\cite{Ellis:1998ct,Tsagas:2007yx}. Any observer can find, say, the unique CMB reference frame, or the energy frame (in which the net heat flux vanishes), and use such a frame to define the electric and magnetic parts of the Weyl tensor which do not depend on any coordinate system, thereby preserving general covariance. That is, there are various velocity fields around which physically and invariantly define the rest of the spacetime quantities; the coordinate system one uses remains irrelevant.  Such considerations have led to the 1+3 covariant approach to cosmological perturbation theory in which gauge-invariance and frame-invariance are subtly separated. 

The 1+3 approach is a semi-tetrad conversion of the field equations of GR into a set of evolution and constraint equations which are derived from the Ricci and Bianchi identities. The variables involved are all covariant objects (i.e., they are tensors), defined through projections with a physical velocity field $u^a$ and its associated spatial projection tensor $h_{ab}=g_{ab}+u_au_b$ and volume element $\varepsilon_{abc}=u^d\eta_{abcd}$~\cite{Ellis:1998ct}.
All tensors may then be irreducibly decomposed into scalars, vectors and projected symmetric and trace-free (PSTF) tensors. These rank-0 tensors, projected rank-1 tensors and PSTF rank-2 tensors  collectively describe the dynamics of the spacetime and are governed by a system of first-order PDEs.  In cosmology, where there is approximate spatial homogeneity and isotropy, such a projection leads to the background spacetime being described by a family of invariant scalars, such as the energy density and expansion rate, provided $u^a$ is the velocity of the fluid. 

More usefully than this is what happens at first-order. Any of the 1+3 objects with an index must vanish in the background as otherwise it would break the symmetry, and so it must be GI by the Stewart-Walker Lemma. This important theorem states that a tensor is GI provided it vanishes in the background~\cite{Stewart:1974uz}. Except in specialised situations little work has been carried out at higher order~\cite{Clarkson:2003af,Clarkson:2011td} because of the difficulty in finding GI quantities.

\subsection{Local scalar-vector-tensor decomposition}

All  rank-1 and -2 tensors used here are
orthogonal to $u^a$, and rank-2 tensors are projected with $h_{ab}$, symmetric and
trace-free which we denote using angle brackets on indices. We define the spatial covariant derivative acting on scalars or spatial
tensors as $\Del_aX_{b\dots c}=h_a^{~a'}h_b^{~b'}\cdots h_c^{~c'}\del_{a'}X_{b'\dots c'}$. The irreducible parts of
the spatial derivative of PSTF tensors are the divergence, curl,
and distortion, defined as~\cite{Maartens:1996ch}
\ba
\div X_{b\dots c}&=& \sdel^a X_{ab\dots c}\\
\curl X_{ab\dots c}&=&\varepsilon_{de\<a}\sdel^d X_{b\dots c\>}^{~~~~~e}\\
\dis X_{ca\dots b}&=&\sdel_{\<c}X_{a\dots b\>}.
\ea
Then, the spatial derivative of a rank-$n$ PSTF tensor $X_{A_n}=X_{a_1a_2\ldots a_n}$ may be decomposed as (for $n=1,2,3$)
\be
\sdel_b X_{A_n}=\frac{2n-1}{2n+1}\,\div X_{\<A_{n-1}}h_{a_n\>b}
-\frac{n}{n+1}\,\curl X_{c\<A_{n-1}}\varepsilon_{a_n\>b}^{~~~~~c}
+\dis X_{bA_n}\,.
\ee 
Note that the divergence decreases the rank of the tensor by one,
the curl preserves it, while the distortion increases it by one (and all are PSTF).
Keeping this in mind one can drop the indices on differential
operators as long as it's explicit the valance of the PSTF tensor
which is being acted on.

At maximal perturbative order a GI object may be considered as a field on an FLRW background. Then, a general rank-2 GI PSTF tensor can be split into a non-local scalar, vector and tensor parts 
\ba
X_{ab}=S_{ab}+V_{ab}+T_{ab}=\sdel_{\<a}\sdel_{b\>}S+\sdel_{\<a}V_{b\>}+T_{ab},\label{SVT}
\ea
where the scalar part $S_{ab}$ is curl-free, the vector part $V_{ab}$ is
solenoidal,  $\sdel^aV_a=0\Rightarrow\sdel^a\sdel^bV_{ab}=\sdel^a\sdel^b\sdel_{\<a}V_{b\>}=0$, while the tensor part is transverse, $\div T_{a}=0$. A similar decomposition exists for rank-1 tensors, but we shall concern ourselves here with rank-2 tensors only: a PSTF rank-2 tensor can be formed from a rank-1 one by taking a distortion without affecting its SVT classification. An important problem using such variables is that they are non-local~\cite{Stewart:1990fm}.

It was demonstrated in~\cite{Clarkson:2011td} that there exists  equivalent \emph{local} variables corresponding to $S_{ab}$, $V_{ab}$ and $T_{ab}$ which can be found by acting with suitable combinations of differential operators on $X_{ab}$. These are local because they do not rely on the solution to an elliptic differential equation that the usual SVT decomposition~(\ref{SVT}) implicitly relies on; consequently there is no reliance on integrals over all space and unknowable boundary conditions.
  These differential operators, which act on rank-2 PSTF tensors, are:\footnote{Note that $\ten$ looks a bit different from in \cite{Clarkson:2011td}. We have substituted for the curvature in terms of $\curl^2$ and $\Del^2$ so that $\ten$ can be used unambiguously as a differential operator in any spacetime.}
\ba
\scal&=&\dis\sdel\div\div \\
\vec&=&\dis\curl\div \\
\ten&=&
\left[\dis\div-\sfrac{1}{3}\Del^2+\sfrac{2}{3}\curl^2\right]\curl\,.
\ea
These are constructed so as to preserve the rank of $X_{ab}$.\footnote{Written out in full these are: 
\ba\fl
\scal X_{ab}=\Del_{\<a}\Del_{b\>}\Del^c\Del^d X_{cd}\\
\fl\vec X_{ab}=\varepsilon_{cd\<a}\D_{b\>}\D^c\D^e{X_e}^d\\
\fl\ten X_{ab}= \sfrac{1}{2}\varepsilon_{cde}\D_{\<a}\D^e\D^c{X_{b\>}}^d 
+\sfrac{1}{6}\varepsilon_{cd\<a}\bigl[ 3\D_{b\>}\D^e\D^c{X_{c}}^d
-\D^2\D^c{X_{b\>}}^d-4\D^c\D^2{X_{b\>}}^d\nonumber\\\fl
~~~~~-\D^e\D^c\D_e{X_{b\>}}^d+\D^e\D^c\D_{b\>}{X_e}^d+\D^c\D^e\D_{b\>}{X_e}^d
+\D^e\D_{b\>}\D^c{X_e}^d+\D^c\D^e\D^dX_{b\>e}
\bigr]
\ea
} 
Note that $\div\ten=0$, $\div\div\vec=0$ and $\curl\scal=0$ as required.  Then, $\scal X_{ab}$ depends only on $S_{ab}$, $\vec X_{ab}$ depends only on $V_{ab}$, $\ten X_{ab}$ depends only on $T_{ab}$; similarly, $\ten X_{ab}$ is a divergence-free rank-2 tensor, hence a bone fide tensor mode. These operators are mutually annihilating when acting on a quantity at maximal perturbative order (i.e., acting on the background) in the sense that 
\be
\scal\vec=\vec\scal=\scal\ten=\ten\scal=\vec\ten=\ten\vec=0\,.
\ee

\section{Gauge-invariant objects}

At first-order finding gauge-invariant objects is trivial: any object with an index must be GI by the Stewart-Walker lemma. How do we find GI objects at second- and higher-order? The trick in going from zeroth to first lies in taking gradients of scalars: scalars are non-zero in the background but their gradients will not be, thereby providing useful GI objects (in addition to things like the electric and magnetic Weyl tensors which are GI anyway). At first-order, however, all of the standard 1+3 covariant quantities are non-zero, so it has been a long standing problem as to how GI second-order objects may be formed. Can we take derivatives of first-order quantities in a suitable way such that they are zero until we get to second-order? What about higher-orders?

One method is to excite only a certain degree of freedom at first-order, such as scalar modes~\cite{Clarkson:2003af,Clarkson:2011td}. Then, any variable which is a pure vector or tensor mode has to be second-order and, hence, gauge-invariant. For example, in this case we can easily see that the vorticity is a pure vector as it is divergence free~-- consequently it must be GI at second-order; similarly, $\vec E_{ab}$ is also a GI vector mode at second-order, and $\ten H_{ab}$ is a GI tensor mode. What is not obvious is how to isolate the scalar modes which are induced at second-order in a GI way. This actually illustrates the main problem: how do we handle the general situation when \emph{all} modes are excited at first?

The key lies in analysing what happens in going from zeroth to first. Consider the gradient of the energy density $\Del_a\rho$, which is GI at first-order. Despite appearances, $\Del_a\rho$ is actually a mixture of a scalar and a vector mode. We can see it contains a vector mode because if we take the curl, we find $\curl\Del_a\rho\propto\omega_a$. Similarly, to isolate the scalar part of $\D_a\rho$ we have to take a divergence. The reason for this is that because $\rho$ itself is not GI when we take its gradient, the covariant derivative is not a background covariant derivative but has first-order connection terms mixing in; the vector degree of freedom arises from this.  To summarise, we must first take a derivative orthogonal to the symmetry of the background to form a GI object; then we must differentiate again in two ways to isolate the scalar and vector part. 

Now let us generalise this to one order higher. Consider $\dis\omega_{ab}$: this is a pure vector degree of freedom at first-order because $\omega_a$ is divergence free. Consequently, at first-order, we must have $\scal\dis\omega_{ab}=0$ and $\ten\dis\omega_{ab}=0$. Therefore, \emph{at second-order $\scal\dis\omega_{ab}$ and $\ten\dis\omega_{ab}$ must be GI PSTF tensors}. Thinking of $\scal$, $\vec$ and $\ten$ as `orthogonal operators', taking derivatives orthogonal to $\vec$ (since $\dis\omega_{ab}$ is a pure vector) results in GI variables at the next perturbative order.  Given either of these objects we can operate once more with $\scal$, $\vec$ or $\ten$ to isolate pure scalar, vector or tensor degrees of freedom. 

More generally, if we take any PSTF rank-2 tensor such as the shear or electric Weyl curvature, we can operate with $\scal$, $\vec$ and $\ten$ to form a set of 3 new PSTF objects: in the case of the shear for example we have $\scal\sigma_{ab}$, $\vec\sigma_{ab}$ and $\ten\sigma_{ab}$. At first-order these have an invariant meaning in terms of the SVT decomposition, but at second they do not. Now operate on each of these three with $\scal$, $\vec$ and $\ten$, to form a total of 9 new rank-2 tensors. Of these, $\scal\scal\sigma_{ab}$, $\vec\vec\sigma_{ab}$ and $\ten\ten\sigma_{ab}$ are not much use to us as they are non-zero at first-order. However, \emph{all the mixed cases have to vanish at first-order and therefore must be GI at second}. That is, an operator to produce a GI PSTF tensor at second-order from any PSTF first-order one belongs to the annihilating set
\be
{\bm \Pi}=\left\{\scal\vec,\scal\ten,\vec\scal,\vec\ten,\ten\scal,\ten\vec\right\}
\ee
After this operation, they \emph{still} do not have an invariant meaning in terms of the SVT split because $\scal$, $\vec$ and $\ten$ must act on GI quantities for this to be the case (analogously to $\D_a\rho$ not being a scalar at first-order). To extract the SVT parts of an element of ${\bm \Pi}[\sigma_{ab}]$, say, we must differentiate again with $\scal$, $\vec$ or $\ten$, giving a total of 6 scalars, 6 vectors and 6 tensors. 
Explicitly, we have:
\ba
\fl\text{scalar:}~~\scal{ \Pi_A}\sigma_{ab} ~\rightsquigarrow~ \scal\scal\vec\sigma_{ab},\scal\scal\ten\sigma_{ab},\scal\vec\scal\sigma_{ab},\scal\vec\ten\sigma_{ab},\scal\ten\scal\sigma_{ab},\scal\ten\vec\sigma_{ab}\nonumber\\
\fl\text{vector:}~~\vec{ \Pi_A}\sigma_{ab} ~\rightsquigarrow~\vec\scal\vec\sigma_{ab},\vec\scal\ten\sigma_{ab},\vec\vec\scal\sigma_{ab},\vec\vec\ten\sigma_{ab},\vec\ten\scal\sigma_{ab},\vec\ten\vec\sigma_{ab}\nonumber\\
\fl\text{tensor:}~~\ten{\Pi_A}\sigma_{ab}~\rightsquigarrow~ \ten\scal\vec\sigma_{ab},\ten\scal\ten\sigma_{ab},\ten\vec\scal\sigma_{ab},\ten\vec\ten\sigma_{ab},\ten\ten\scal\sigma_{ab},\ten\ten\vec\sigma_{ab},\nonumber
\ea 
where $A=1,\ldots,6$, forming a set of 18 GI objects which now have an invariant meaning in terms of the SVT split. Similar quantities can be formed starting from any PSTF rank-2 tensor, including things like $\Del_{\<a}\Del_{b\>}\rho$. 

Let us now extend this to third-order. Again we can form objects which vanish at second-order by applying  annihilating projections of $\scal$, $\vec$ and $\ten$ to those SVT GI variables just defined~-- i.e., $\scal\vec{ \Pi_A},\scal\ten{ \Pi_A},\vec\scal{ \Pi_A},\vec\ten{ \Pi_A},\ten\scal{ \Pi_A},\ten\vec{ \Pi_A}$ give a total of $6\times6=36$ objects which vanish at second-order and consequently must be GI at third. Note that we have essentially acted with $\bm\Pi$ twice on a first-order PSTF tensor in order to find all these objects. Further action with $\scal$, $\vec$ or $\ten$ will then split these into sets of scalars, vectors and tensors. Thus, $\ten\vec\scal\vec\scal\sigma_{ab}$ is a GI third-order tensor mode.  

We can generalise this to any order. We have seen that at second-order $\bm\Pi$ acting on a first-order PSTF tensor gives a set of 6 GI variables, while  at third we have to act with $\bm\Pi\bm\otimes\bm\Pi$, i.e., the outer product of $\bm\Pi$ with itself. In index notation we have $\Pi_A\Pi_B$. At perturbative order $n$ we must act with $\bm\Pi$ $n-1$ times, or 
\be
\bm\Pi^{\otimes(n-1)}=\underbrace{\bm\Pi\bm\otimes\bm\Pi\bm\otimes\cdots\bm\otimes\bm\Pi}_{n-1~~\mbox{times}}\,.
\ee
In index notation we have instead $\Pi_{A_1}\Pi_{A_2}\cdots\Pi_{A_{n-1}}$, giving a total of $6^{n-1}$ possible GI PSTF tensors. The SVT parts can be separated with a final action of $\scal$, $\vec$ and $\ten$ as before.

\section{Conclusions}

We have demonstrated for the first time how to generate covariant, locally defined, GI objects at any order when perturbing away from an FLRW background. Utilising local differential operators which  mutually annihilate on the background we can form rank-2 tensors which represent only scalar, vector or tensor degrees of freedom, provided they act upon a GI object at a given perturbative order. Recursively acting with these operators in the way described above then provides a neat way to define a large variety of GI variables at any given order.  

How would such GI quantities be used in practise? In principle, for a given order $n$ one can calculate  evolution equations for $\bm\Pi^{\otimes(n-1)}\bm X_{ab}$, where $\bm X_{ab}$ stands for a set of appropriate rank-2 tensors, using standard commutation relations and the 1+3 covariant equations. Products of quantities which appear will be of perturbative order $n-1$ and can be written in terms of  products of $\bm\Pi^{\otimes(n-2)}\bm X_{ab}$; these will have their own evolution equations. The final system will yield a large system of PDEs which can be solved order by order. This is left for future work. 

The physical information these variables contain will be very subtle to interpret of course. When considering something like $\ten\scal\vec\sigma_{ab}$ for example, we cannot read this as the `the tensor part of the scalar part of the vector part of the shear at second-order', because once one moves up a perturbative order the meaning of the SVT decomposition disappears. Analogously, $\vec\sigma_{ab}$ is the vector part of the shear at first-order, but not at second. But such difficulties are to be expected: there are only a limited number of tensors that are regularly used in practise which we have intuition for, so to expect one to `appear' at third- or tenth-order representing a GI tensor mode which \emph{also} has a simple physical interpretation is implausible. 
Nevertheless, as higher-order perturbation theory is developed in full generality this will be an important avenue to explore.

\ack I would like to thank Marco Bruni, George Ellis, Kenneth Hughes, Roy Maartens, Bob Osano and Obinna Umeh for comments and/or discussions.

\end{document}